\documentclass[aps,showpacs,twocolumn,prl,superscriptaddress]{revtex4}

\usepackage[tbtags]{amsmath}
\usepackage{amssymb}
\usepackage{bmpsize}
\usepackage{mathtools}
\usepackage{caption}

\usepackage{amsfonts}    % need for subequations
\usepackage{graphicx}   % need for figures
\usepackage{verbatim}   % useful for program listings
\usepackage{color}      % use if color is used in text
\usepackage{subfigure}  % use for side-by-side figures
\usepackage{hyperref}   % use for hypertext links, including those to external documents and URLs
\usepackage{natbib}
\usepackage{booktabs}
\usepackage{threeparttable}
\usepackage{dcolumn}
\usepackage{multirow}
\usepackage{fancyhdr}

\newcommand{\red}[1]{\textcolor{black}{#1}}
\newcommand{\blue}[1]{\textcolor{black}{#1}}

\begin{document}
\title{Observation of nonlocality sharing via not-so-weak measurements}
\author{Tianfeng Feng}
%\altaffiliation{These authors contributed equally}
\affiliation{State Key Laboratory of Optoelectronic Materials and Technologies and School of Physics, Sun Yat-sen University, Guangzhou, People's Republic of China}
 
\author{Changliang Ren}
\email{renchangliang@cigit.ac.cn}
\affiliation{Center for Nanofabrication and System Integration, Chongqing Institute of Green and Intelligent Technology, Chinese Academy of Sciences, Chongqing, People's Republic of China}
\affiliation{School of Physics and Electronics, Hunan Normal University, Changsha 410081, Hunan, People's Republic of China}

\author{Yuling Tian}
%\altaffiliation{These authors contributed equally}
\affiliation{State Key Laboratory of Optoelectronic Materials and Technologies and School of Physics, Sun Yat-sen University, Guangzhou, People's Republic of China}

\author{Maolin Luo}
\affiliation{State Key Laboratory of Optoelectronic Materials and Technologies and School of Physics, Sun Yat-sen University, Guangzhou, People's Republic of China}

 \author{Haofei Shi}
\affiliation{Center for Nanofabrication and System Integration, Chongqing Institute of Green and Intelligent Technology, Chinese Academy of Sciences, Chongqing, People's Republic of China}

\author{Jingling Chen}%\email{Correspondence to: chenjl@nankai.edu.cn}
 \affiliation{Theoretical Physics Division, Chern Institute of Mathematics, Nankai University,
 Tianjin 300071, People's Republic of China}
 \affiliation{Centre for Quantum Technologies, National University of Singapore,
 3 Science Drive 2, Singapore 117543}

 \author{Xiaoqi Zhou}\email{zhouxq8@mail.sysu.edu.cn}
\affiliation{State Key Laboratory of Optoelectronic Materials and Technologies and School of Physics, Sun Yat-sen University, Guangzhou, People's Republic of China}
\begin{abstract}

Nonlocality plays a fundamental role in quantum information science. Recently, it has been theoretically predicted and experimentally demonstrated that the nonlocality of an entangled pair may be shared among multiple observers using weak measurements with moderate strength. Here we devise an optimal protocol of nonlocality sharing among three observers and show experimentally that nonlocality sharing may be also achieved using weak measurements with near-maximum strength. Our result sheds light on the interplay between nonlocality and quantum measurements and, may find applications in quantum steering, unbounded randomness certification and quantum communication network.

%\rd{We have also designed a quantum game whose optimal strategy is dependent on the nonlocality sharing protocol.}
%\blue{Besides, we designed a quantum game in which the optimal strategy of three players for winning the game corresponds to the case of optimal nonlocality sharing.}

%We have shown that non-locality sharing can be achieved via ``strong'' weak measurements. An experiment has been reported. We have shown that one can achived a double Bell inequality violation among three observers with a weak measurement as strong as possible.
\end{abstract}

%================================================================================

\maketitle

%Nonlocality, the phenomenon that the results of local \red{ measurements} performed on distant parties of a composite system can not be explained by local hidden variable theories, is one of the most  fascinating characteristics of quantum mechanics\cite{Brunner}. 

\section{Introduction} The correlations observed among the results of measurements performed on distant parties of a composite quantum system cannot be explained by local hidden variable theories. This is the phenomenon of nonlocality and represents  one of the most fascinating features of quantum mechanics~\cite{Brunner}. Since the derivation of Bell inequalities in 1964 \cite{Bell}, nonlocality has been studied extensively from various perspectives~\cite{Clauser,Zukowski,Mermin,Belinskii,Ardehali,Collins,Brukner, Lee} and verified in many different quantum systems~\cite{Aspect,Weihs,Rowe,Hofmann,Giustina,Christensen,Hensen,Giustina1,Shalm}. Nonlocality is crucial for our understanding of quantum mechanics, and represents a resource in device-independent quantum information protocols, such as quantum key distribution~\cite{Acin}, randomness generation~\cite{Pironio, Liu, Bierhorst}, and entanglement certification~\cite{Bowles}.

%A typical scenario of nonlocality is manifested by Bell inequality. Alice and Bob perform measurements on their respective particles. Alice and Bob perform different measurements. This is the so-called Bell inequality. 
As a matter of fact, nonlocality has mostly discussed for pairs of entangled qubits distributed to two separated observers. Recently, a surprising result has been reported \cite{Silva}, showing that nonlocality of a pair of qubits may be actually shared among more than two observers using weak measurements. In this scenario, a pair of maximally-entangled qubits are distributed to three observers Alice, Bob1 and Bob2, with Alice having accesse to one qubit and the two Bobs to the other. At first, Bob1 performs a weak measurement on his qubit and then passes it to Bob2 and then, Alice and Bob2 perform projective measurements on their own qubits respectively. Nonlocality sharing, which is certified by a double violation of  \red{Bell-CHSH} inequality among Alice-Bob1 and Alice-Bob2, may be observed when the strength of Bob1's measurement is within a reasonable range \cite{Silva, Hu,Schiavon}. This constraint can be understood as follows: The \red{strength} of Bob1's measurement has to be large enough to \red{retrieve} enough information from the qubit and to establish quantum correlation between Alice and Bob1,  but not too large, for Bob2 to retain some quantum correlation with Alice.

\begin{figure}[htbp]
\includegraphics[width=0.5\textwidth]{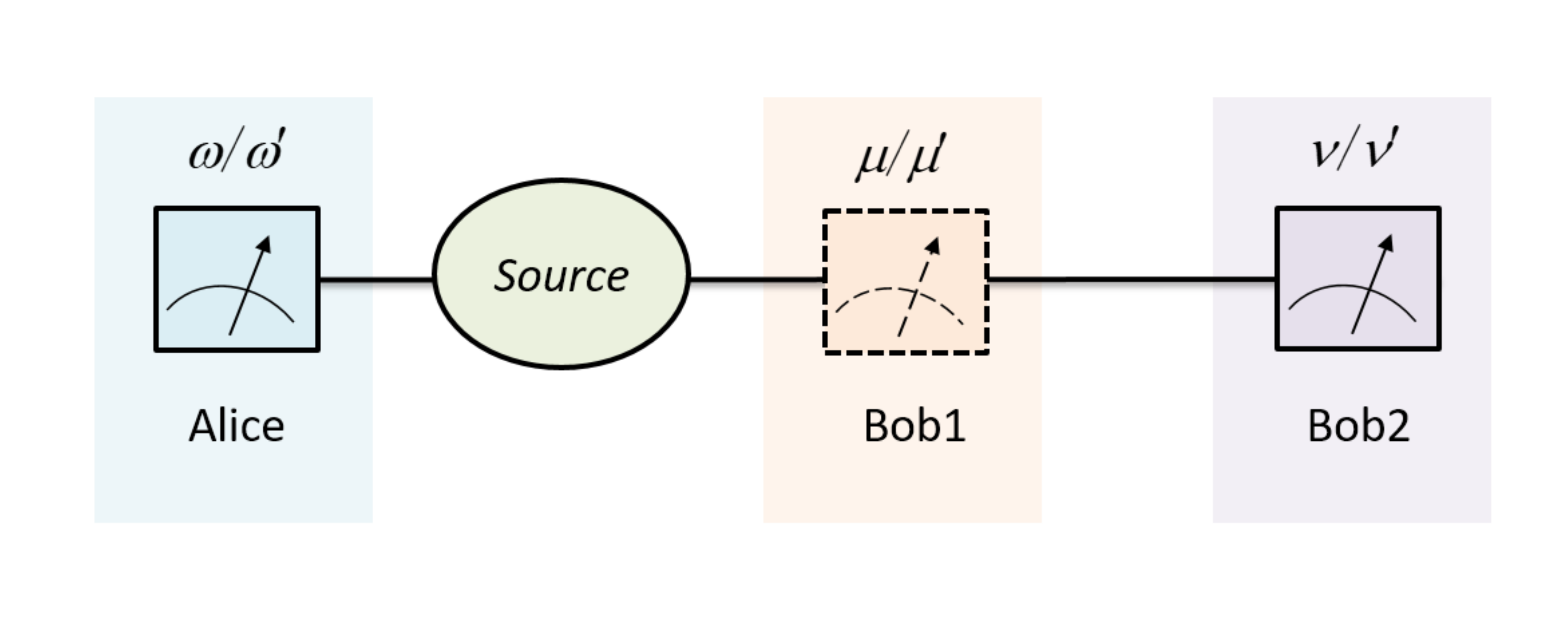}
\caption{Schematic diagram of nonlocality sharing. A two-qubit entangled state is distributed to three observers, Alice, Bob1 and Bob2, where Bob1 and Bob2 have access to the same qubit. Alice and Bob2 implement projective measurements on their qubits, whereas Bob1 implements weak measurement on his qubit.
}
\label{setup}
\end{figure}
%, in which the strength of the measurement is strong enough to gain quantum correlation between Alice and Bob1 , is weak enough to retain quantum correlaion between Alice and Bob2 

Recently, much attention has been paid to nonlocality sharing \cite{DAS,Mal,Ren}. %Especially, Ren \emph{et al} extended Sliva's scheme and proved a counterintuitive result that nonlocality sharing can actually be achieved even with almost strong measurements\cite{Ren}. 
However, all proposed protocols are not optimal, i.e. they do not maximize the violation of Bell inequalites at different strengths of the weak measurements. %, and they are not conducive and friendly to experimental verification. 
\blue{In this article, we devise an optimal protocol for nonlocality sharing among three observers where the strength of the measurement performed by Bob1 can be increased to near maximum.} In particular, we report the experimental observation of nonlocality sharing in a photonic system using optimal weak measurements.  In our experiment, we observed a double violation of Bell-CHSH inequality in a region of parameter not included in the original scheme \cite{Silva, Hu,Schiavon} of nonlocality sharing.

\section{Optimal protocol of nonlocality sharing} Let us first briefly review Silva's scheme that allows double violation of Bell-CHSH inequality \cite{Silva}. As shown in Fig.~1, the three observers, Alice, Bob1 and Bob2, perform measurements on a two-qubit entangled state. Alice measures qubit 1 and the two Bobs measure qubit 2 sequentially. Each observer randomly chooses one of two observables to measure. The two observables for Alice~(Bob1/Bob2) are $\hat{\omega}$~($\hat{\mu}$/$\hat{\nu}$) and $\hat{\omega'}$~($\hat{\mu'}$/$\hat{\nu'}$). Alice and Bob1 use their measurement results to construct a CHSH parameter $I^{(1)}_{\mathrm{CHSH}}$, which is defined as
 \begin{eqnarray}\label{CHSH1}
 I^{(1)}_{\mathrm{CHSH}}=\langle \hat{\omega}\otimes\hat{\mu}\rangle+\langle \hat{\omega'}\otimes\hat{\mu}\rangle+\langle \hat{\omega}\otimes\hat{\mu'}\rangle-\langle \hat{\omega'}\otimes\hat{\mu'}\rangle
 \end{eqnarray}
In the same way, Alice and Bob2 construct another CHSH parameter $I^{(2)}_{\mathrm{CHSH}}$ defined as
 \begin{eqnarray}\label{CHSH2}
 I^{(2)}_{\mathrm{CHSH}}=\langle \hat{\omega}\otimes\hat{\nu}\rangle+\langle \hat{\omega'}\otimes\hat{\nu}\rangle+\langle \hat{\omega}\otimes\hat{\nu'}\rangle-\langle \hat{\omega'}\otimes\hat{\nu'}\rangle
  \end{eqnarray}

A CHSH parameter greater than $2$ cannot be explained by any local hidden variable theory. According to the monogamy property of nonlocality \cite{Masanes,Toner}, if the the measurements performed by the three observers are projective, it is impossible to achieve a double \red{Bell-CHSH} violation, namely to have both $I^{(1)}_{\mathrm{CHSH}}$ and $I^{(2)}_{\mathrm{CHSH}}$ greater than 2. \blue{However, if Bob1's measurement is a weak measurement, this restriction no longer holds.} As illustrated in Fig.2(a), by setting $\hat{\omega}=X$, $\hat{\omega'}=Z$, $\hat{\mu}=\hat{\nu}=\frac{X+Z}{\sqrt{2}}$, $\hat{\mu'}=\hat{\nu'}=\frac{X-Z}{\sqrt{2}}$, one obtains
\begin{equation}
I^{(1)}_{\mathrm{CHSH}} =2\sqrt{2}G, I^{(2)}_{\mathrm{CHSH}} =\sqrt{2}(1+F),
%=2\sqrt{2}\cos2\theta \\%=\sqrt{2}(1+\sin2\theta).
\end{equation}
 %(See more details in Methods).
where $G$ is the precision factor of Bob1, i.e. the strength of his weak measurement, quantifying how much information is gained through such measurement, and $F$ is the quality factor, i.e. the coherence left after the measurement. There is a trade-off between $G$ and $F$, which may be written as $G^{2}+F^{2}\leq1$. % and an optimal weak measurement is achieved when the condition $G^{2}+F^{2}=1$ is satisfied~\cite{silva}. 
For an optimal weak measurement, namely $G^2+F^2=1$, the two CHSH parameters as functions of $G$ are shown in Fig.~2(c), where $I^{(1)}_{\mathrm{CHSH}}$ ~($I^{(2)}_{\mathrm{CHSH}}$) is represented by the blue line~(red curve). One can see that, only when the strength of the weak measurement is moderate, namely $G\in(0.707,0.910)$, a double \red{Bell-CHSH} violation may be achieved. 

In Ref. \cite{Silva} it was suggested that a double  \red{Bell-CHSH} violation may be achieved only with a moderate-strength weak \red{measurement}. The line of reasoning was the following: If  Bob1's measurement is too weak, he cannot gain enough information about qubit 2 to make $I^{(1)}_{\mathrm{CHSH}}>2$. If it is too strong, it degrades too much the quantum correlation between the two qubits, which are then not strong enough to make $I^{(2)}_{\mathrm{CHSH}}>2$. \blue{In this article, we point out that such argument is not valid in general, and a double Bell-CHSH violation may actually be achieved using weak measurements with near-maximum strength}.
%with a very ``strong'' measurement as long as it is still a weak measurement.

\begin{figure*}[htbp]
\includegraphics[width=\textwidth]{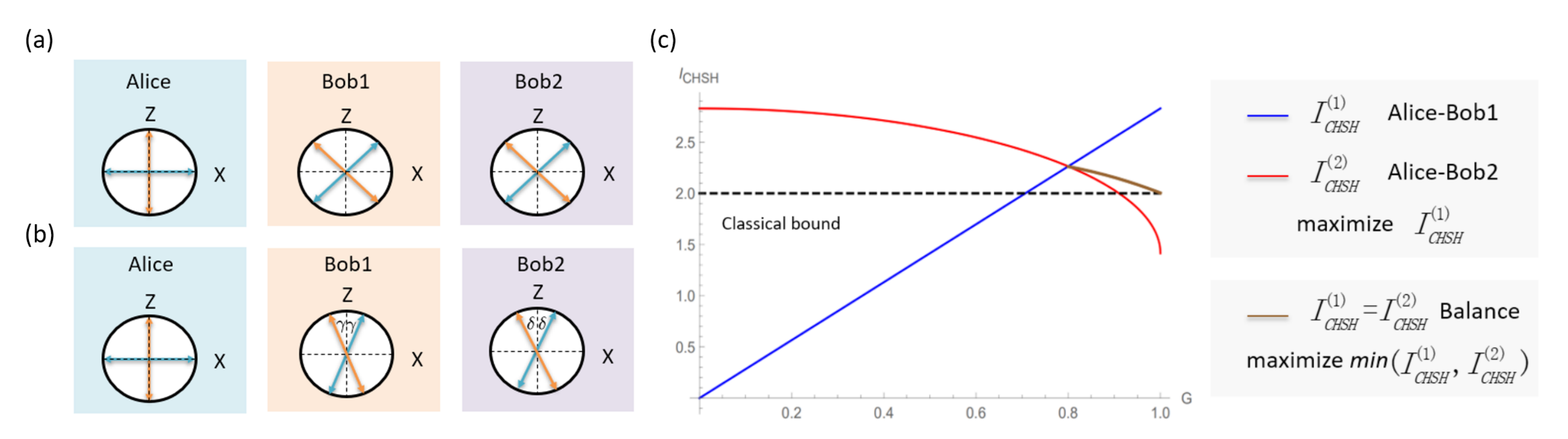}
\caption{ Left: the observables considered in the original nonlocality sharing scheme~(a) and in the present one~(b). Right: the CHSH parameters as a function of Bob1's precision factor G. Blue~(red) line denotes $I^{(1)}_{\mathrm{CHSH}}$~($I^{(2)}_{\mathrm{CHSH}}$) using the~(a) observables. Brown curve corresponds to both $I^{(1)}_{\mathrm{CHSH}}$ and $I^{(2)}_{\mathrm{CHSH}}$ using the new observables under the condition $G>0.8$.
}
\label{setup}
\end{figure*}

Let us investigate the original protocol and explain why it can be improved. In Silva's scheme, Alice's observables are $X$ and $Z$. Based on Alice's observables, Bob1 set his observables to $\hat{\mu}=\frac{X+Z}{\sqrt{2}}$ and $\hat{\mu'}=\frac{X-Z}{\sqrt{2}}$, which lead to the maximum value $I^{(1)}_{\mathrm{CHSH}}=2\sqrt{2}G$. 
However, the choice of Bob1's observables only considers the maximization of $I^{(1)}_{\mathrm{CHSH}}$, which will unavoidably lower the upper bound for $I^{(2)}_{\mathrm{CHSH}}$ \cite{Ren}. To achieve a double \red{Bell-CHSH} violation, one needs to make sure that both CHSH parameters are greater than 2. As a result, to make the double \red{Bell-CHSH} violation region as large as possible, a suitable choice is to maximize the value of $min(I^{(1)}_{\mathrm{CHSH}},I^{(2)}_{\mathrm{CHSH}})$ instead of  $I^{(1)}_{\mathrm{CHSH}}$ or $I^{(2)}_{\mathrm{CHSH}}$ alone.

We now present our scheme of nonlocality sharing. In our protocol, Alice uses the same fixed observables $X$ and $Z$. The choice of Bob1's and Bob2's observables depends on the precision factor $G$ of Bob1's measurement. 

When $G\le0.8$, $I^{(1)}_{\mathrm{CHSH}}$ is always less than or equal to $I^{(2)}_{\mathrm{CHSH}}$ and thus Bob1 and Bob2 may use the original scheme \cite{Silva} to maximize $min(I^{(1)}_{\mathrm{CHSH}},I^{(2)}_{\mathrm{CHSH}})$. 

%When $G\le0.8$, $I^{(1)}_{\mathrm{CHSH}}$ is always less than or equal to $I^{(2)}_{\mathrm{CHSH}}$, which means $min(I^{(1)}_{\mathrm{CHSH}},I^{(2)}_{\mathrm{CHSH}})=I^{(1)}_{\mathrm{CHSH}}$ under the condition of $G\le0.8$.
%Since Silva's scheme can maximize the value of $I^{(1)}_{\mathrm{CHSH}}$, we will just use the same observables as Silva's scheme when $G\le0.8$, because this can maximize the value of $min(I^{(1)}_{\mathrm{CHSH}},I^{(2)}_{\mathrm{CHSH}})$ under the condition of $G\le0.8$.

%We now explain the reasoning behind our choice of Bob1's and Bob2's observables.

When $G>0.8$, $I^{(2)}_{\mathrm{CHSH}}$ is not necessarily greater than $I^{(1)}_{\mathrm{CHSH}}$. 
In this case, to maximize $min(I^{(1)}_{\mathrm{CHSH}},I^{(2)}_{\mathrm{CHSH}})$, one need to increase the value of $I^{(2)}_{\mathrm{CHSH}}$, namely raising the quantum correlation between Alice and Bob2. 
We find that, given a certain $G$ value, the smaller is the difference between Bob1's two observables $\hat{\mu}$ and $\hat{\mu'}$, the larger are the quantum correlation between Alice and Bob2 after Bob1's measurement. 
Following this reasoning, a promising way to increase the value of $min(I^{(1)}_{\mathrm{CHSH}},I^{(2)}_{\mathrm{CHSH}})$ would be changing $\hat{\mu}$ and $\hat{\mu'}$ and making them more similar to each other.
As a result, we set Bob1's and Bob2's observables~(as shown in Fig.~2(b)) as follows
\begin{equation}
\begin{split}
& \hat{\mu}=\cos\gamma X+\sin\gamma Z,~~\hat{\mu'}=\cos\gamma X-\sin\gamma Z,\\
& \hat{\nu}=\cos\delta X+\sin\delta Z,~~\hat{\nu'}=\cos\delta X-\sin\delta Z,
%=2\sqrt{2}\cos2\theta \\%=\sqrt{2}(1+\sin2\theta).
\end{split}
\end{equation}
where $\gamma$ and $\delta$ are angles between 0 and $\pi/4$, whose values is determined by $G$.
One can see that, while the value of $I^{(1)}_{\mathrm{CHSH}}$ is a function of the two parameters $G$ and $\gamma$, the value of $I^{(2)}_{\mathrm{CHSH}}$ is a function of the three parameters $G$, $\gamma$ and $\delta$. Given a certain $G$, once $\gamma$ is set, not only $I^{(1)}_{\mathrm{CHSH}}$ can be calculated, but also a proper value of $\delta$ may be chosen to maximize the value of $I^{(2)}_{\mathrm{CHSH}}$. As a result, given a certain $G$ greater than 0.8, if we start with $\gamma=\pi/4$ and then gradually decrease its value, the value of $I^{(1)}_{\mathrm{CHSH}}$ will decrease, and in the meantime the value of $I^{(2)}_{\mathrm{CHSH}}$ will increase assuming $\delta$ is chosen to maximize $I^{(2)}_{\mathrm{CHSH}}$. By decreasing the value of $\gamma$, the value of $min(I^{(1)}_{\mathrm{CHSH}},I^{(2)}_{\mathrm{CHSH}})$ can be continuously increased until we achieve the condition $I^{(1)}_{\mathrm{CHSH}}=I^{(2)}_{\mathrm{CHSH}}$. 
Notice that for any given G greater than 0.8, we can always find values of $\gamma$ and $\delta$ to achieve that $I^{(1)}_{\mathrm{CHSH}}=I^{(2)}_{\mathrm{CHSH}}$ and thus obtain the maximum value of $min(I^{(1)}_{\mathrm{CHSH}},I^{(2)}_{\mathrm{CHSH}})$.

As shown in Fig.~2(c), the brown curve indicates the values of both $I^{(1)}_{\mathrm{CHSH}}$ and $I^{(2)}_{\mathrm{CHSH}}$ in the region of $G>0.8$. One can see that the two CHSH parameters are always greater than 2 when $0.707<G<1$. As a result, our protocol has expanded the region of double \red{Bell-CHSH} violation  and can realize nonlocality sharing when $0.910<G<1$, which is unattainable in the original scheme. 

\begin{figure*}[htbp]
\includegraphics[width=0.98\textwidth]{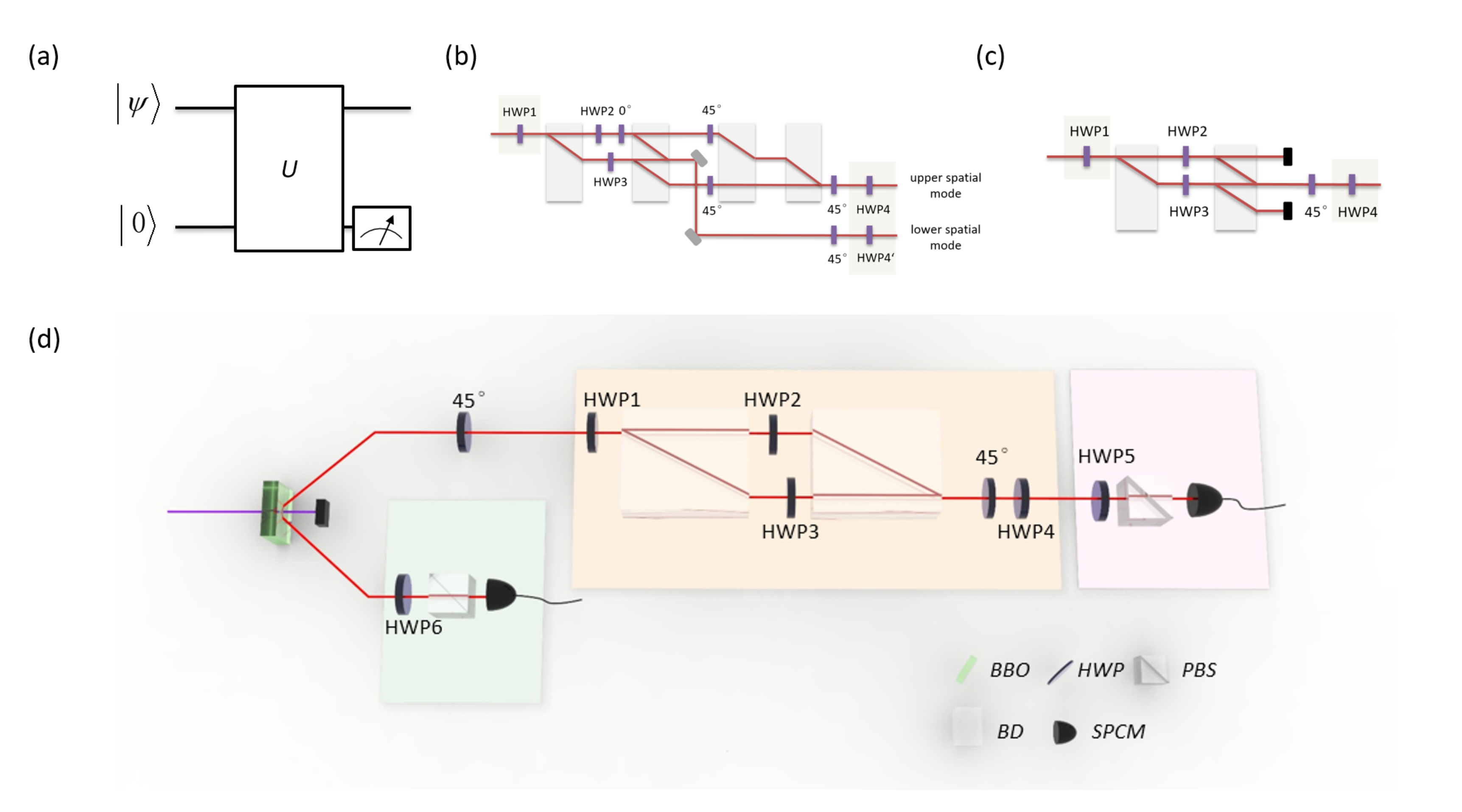}
\caption{Experimental implementation of optimal nonlocality sharing. 
(a) The circuit of optimal weak measurement using an ancillary qubit. 
(b) Optimal weak measurement realized in a photonic system.  
(c) The simplified optical setup for optimal weak measurement. 
(d) Experimental setup.
Degenerated polarization-entangled photon pairs at 808nm are produced by pumping a type-II beta barium borate~(BBO) crystal with an ultraviolet laser centered at 404nm. 
HWP6~(HWP5) is used by Alice~(Bob2) to set the observable of her~(his) projective measurement.
The orange region denotes Bob1's weak measurement setup.
HWP1 and HWP4 are rotated by the same angle and set the observable of Bob1's weak measurement.
HWP2 and HWP3, set at $\pi/2-\theta/2$ and $\theta/2$ respectively, are used to tune precision factor $G$ of Bob1's measurement to $\cos2\theta$.
}
\label{setup}
\end{figure*}

\section{Experimental demonstration of optimal nonlocality sharing} The key to experimentally demonstrate our scheme is to realize a weak measurement with tunable strength. Let us first explain how to realize a weak measurement using an ancillary qubit.
%In the case of the ancillary-qubit assisted weak measurement described above, $G=\cos{2\theta}$ and $F=\sin{2\theta}$, which satisfies the optimal weak measurement condition.
%We first review how to realize an optimal weak measurment with an ancillary qubit. 
As shown in Fig.~3(a), an arbitrary single qubit state $\vert\psi\rangle$, defined as $\vert\psi\rangle=\alpha\vert0\rangle+\beta\vert1\rangle$~($|\alpha|^{2}+|\beta|^{2}=1$), is to be weak-measured. The qubit $\vert\psi\rangle$  is coupled to an ancillary qubit prepared in the state $\vert0\rangle$ by a two-qubit unitary $U$ defined as
\begin{equation}
U=\left(
      \begin{array}{cccc}
        \cos\theta \ & -\sin\theta \ & 0 \ & 0 \\
        \sin\theta \ & \cos\theta \ & 0 \ & 0  \\
        0 \ &  0  \  &\sin\theta  \  &-\cos\theta\\
	0 \ &  0  \  &\cos\theta  \  &\sin\theta\\
      \end{array}
    \right).
\end{equation}
The resulting two-qubit state is given
\begin{equation}
\alpha|0\rangle\otimes(\cos\theta|0\rangle+\sin\theta|1\rangle)+\beta|1\rangle\otimes(\sin\theta|0\rangle+\cos\theta|1\rangle).
\end{equation} 
A projective measurement on $0/1$ basis is then performed on the ancillary qubit, which would effectively realize a weak \red{measurement} on $\vert\psi\rangle$. By easy calculation, one can find $G=\cos{2\theta}$ and $F=\sin{2\theta}$, which satisfies the optimal weak measurement condition \cite{Hu}.

We now show how to experimentally realize a weak measurement on a polarization-encoded photonic qubit in a specific basis $\{\vert \varphi\rangle, \vert \varphi^{\perp}\rangle\}$, where $\vert \varphi\rangle=\cos\varphi\vert H\rangle+\sin\varphi\vert V\rangle$ and $\vert \varphi^{\perp}\rangle=\sin\varphi\vert H\rangle-\cos\varphi\vert V\rangle$~($H/V$ denotes horizontal/vertical polarization). As shown in Fig.~3(b), the optical weak measurement setup consists of three parts. The first part is a half-wave plate~(HWP1) set at $\varphi/2$, which is used to transform the measurement basis from  $\{\vert \varphi\rangle, \vert \varphi^{\perp}\rangle\}$ to $\{\vert H\rangle, \vert V\rangle\}$. The second part, composed of the devices between HWP1 and HWP4, HWP4', is used to implement a weak measurement in $\{\vert H\rangle, \vert V\rangle\}$ basis. The third part, consisting of HWP4 and HWP4' both set at  $\varphi/2$, is used to transform the measurement basis from  $\{\vert H\rangle, \vert V\rangle\}$ back to $\{\vert \varphi\rangle, \vert \varphi^{\perp}\rangle\}$.

The essential part of this setup is the second part, which realizes a weak measurement in the $\{\vert H\rangle, \vert V\rangle\}$ basis by exploiting the spatial degree of freedom as the ancillary qubit. Suppose that the qubit entering the second part is $\vert\psi\rangle=\alpha\vert H\rangle+\beta\vert V\rangle$. By setting HWP2 and HWP3 at $\theta/2$, the state outgoing from part 2 will become 
\begin{equation}
\alpha\vert H\rangle(\cos\theta \vert l\rangle+\sin\theta \vert u\rangle)+\beta\vert V\rangle(\sin\theta \vert l\rangle+\cos\theta \vert u\rangle),
\end{equation} 
where $\vert l\rangle$ and $\vert u\rangle$ denote the lower and upper spatial modes, respectively~(See Methods). This state has the same form as the one shown in equation 6, and thus may be used to realize an optimal weak measurement by measuring the spatial qubit.% one can realize an optimal weak measurement on the polarization qubit in $H/V$ basis with a precision factor $G=\cos2\theta$. 

Actually, one may use a simplified setup to realize the same optimal weak measurement \cite{Hu}. This is illustrated in Fig.~3(c): when the input photon is found in central spatial mode, it indicates that an optimal weak measurement in the basis $\{\vert \varphi\rangle, \vert \varphi^{\perp}\rangle\}$ with +1 outcome has been implemented. By changing the angles of HWP1 and HWP4 from $\varphi/2$ to $\varphi/2+\pi/4$, the same setup effectively implements an optimal weak measurement with -1 outcome. %Note that measurement outcome values are extracted in the final coincidence detection

\begin{figure}[b]
\begin{center}
\includegraphics[width=0.48\textwidth]{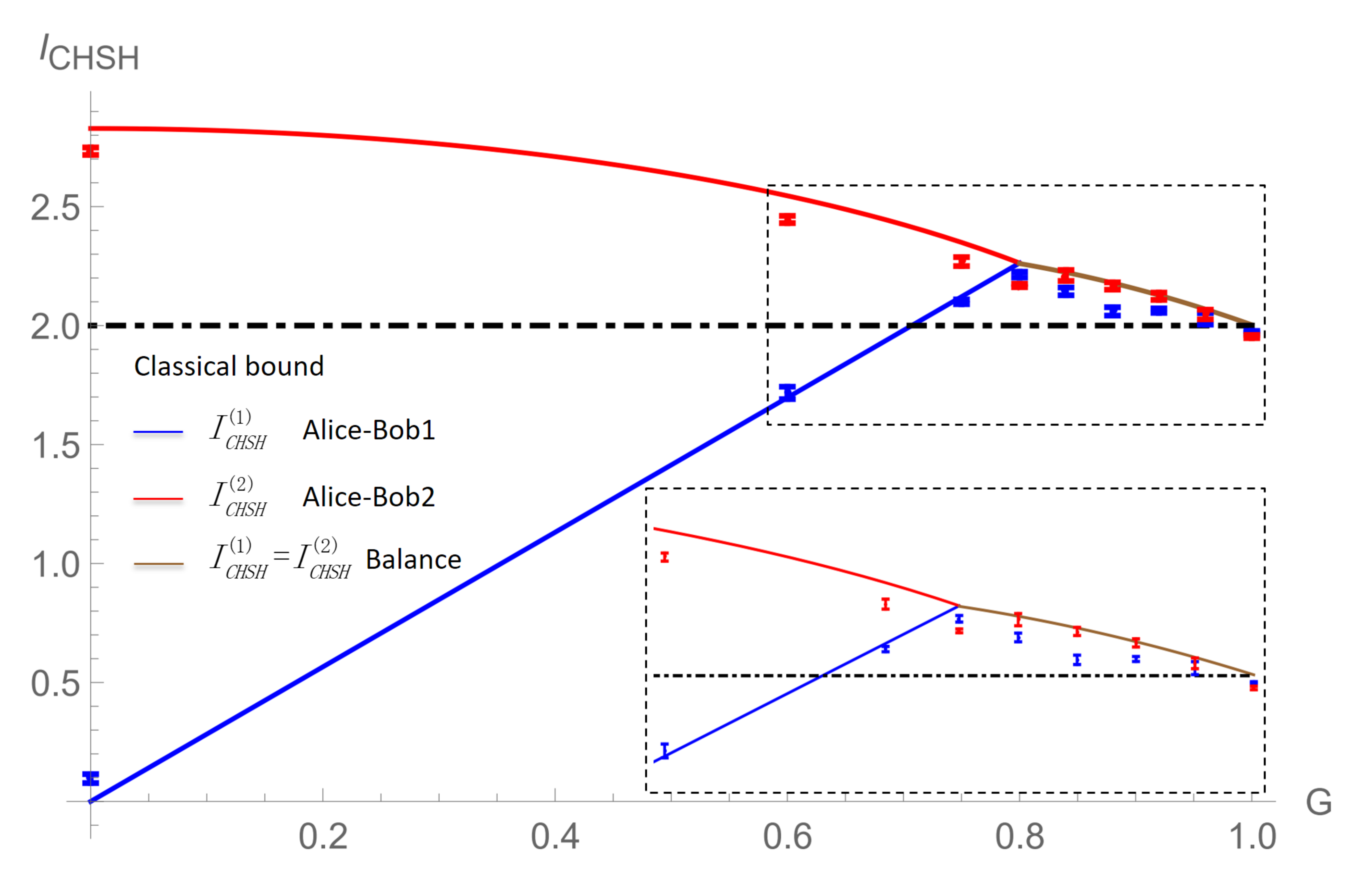}
\end{center}
\par
%It can be used to
%construct GHZ-type argument.
\caption{Experimental results of optimal nonlocality sharing. 
The blue and red curves denote the theoretical predictions for $I^{(1)}_{\mathrm{CHSH}}$ and $I^{(2)}_{\mathrm{CHSH}}$ when $G\le 0.8$, respectively. The brown curve represents the theoretical predictions for both $I^{(1)}_{\mathrm{CHSH}}$ and $I^{(2)}_{\mathrm{CHSH}}$ when $G>0.8$.
The blue and red dots indicate experimental results for $I^{(1)}_{\mathrm{CHSH}}$ and $I^{(2)}_{\mathrm{CHSH}}$ when $G = \{0, 0.6, 0.75, 0.8, 0.84, 0.88, 0.92, 0.96, 1\}$.
The error bars (SD) have been calculated assuming Poissonian counting statistics.}
\end{figure}

 After explaining how to realize an optimal weak measurement on a polarization-encoded qubit, we now report our experiment of nonlocality sharing in a photonic system. As shown in Fig.~3(d), a standard type-II spontaneous parametric down-conversion source is used to produce a two-photon maximally entangled state $\frac{1}{\sqrt{2}}(\vert H_1\rangle\vert H_2\rangle-\vert V_2\rangle\vert V_2\rangle)$, where photon 1 is that of Alice and photon 2 is sent to Bob1 and Bob2. On Alice's side, photon 1 pass through HWP6 and PBS1, which together implement a projective measurement of  $\omega=X$ or $\omega'=Z$. Bob1 lets photon 2 go through the weak measurement setup~(the orange region), which implements an optimal weak measurement of  $\mu=\cos\gamma X+\sin\gamma Z$ or $\mu'=\cos\gamma X-\sin\gamma Z$ with a precision factor $G$, where $\gamma$ is controlled by HWP1 and HWP4, and $G$ by HWP2 and HWP3. After the weak measurement, photon 2 is then passed to Bob2, who implements a projective measurement of $\nu=\cos\delta X+\sin\delta Z$ or $\nu'=\cos\delta X-\sin\delta Z$ by tuning HWP5 and PBS2, where $\delta$ is controlled by the angle of HWP5. Both $\gamma$ and $\delta$ are properly chosen to achieve the precision factor $G$ and to \red{maximize} $min(I^{(1)}_{\mathrm{CHSH}},I^{(2)}_{\mathrm{CHSH}})$ given a certain $G$ value.

In our experiment, we have chosen 9 different values of $G$ and set the measurement bases accordingly. The two CHSH parameters $I^{(1)}_{\mathrm{CHSH}}$ and $I^{(2)}_{\mathrm{CHSH}}$, which are constructed using the \red{measurement} results, are shown in Fig.~4. Theoretically, $min(I^{(1)}_{\mathrm{CHSH}},I^{(2)}_{\mathrm{CHSH}})$ reaches its maximal value 2.263 when $G=0.8$. Experimentally, we get $I^{(1)}_{\mathrm{CHSH}}=2.214\pm0.011$ and $I^{(2)}_{\mathrm{CHSH}}=2.168\pm0.007$, which are both more than 10 standard deviations above the classical bound. Even when the strength of Bob1's measurement reaches a vey high level, say $G=0.96$, we can still observe a double \red{Bell-CHSH} violation $I^{(1)}_{\mathrm{CHSH}}=2.028\pm0.024$ and $I^{(2)}_{\mathrm{CHSH}}=2.047\pm0.020$, which would be impossible using the original protocol. In general, our experimental results are in good agreement with the theoretical predictions. The mismatch with the theoretical preditions is mainly due to the imperfections of the photon source and losses in the interferometeric setup.

\section{conclusions} In summary, we have devised an optimal protocol to realize nonlocality sharing among three observers and experimentally demonstrated it using weak measurements with a wide range of measurement strength.
\blue{Unlike previous experiments, where double violations of Bell-CHSH inequality were only be observed using a weak measurement with moderate strength, we have demonstrated that double violations may be achieved via weak measurements with near-maximum strength. } %\blue{Besides, we designed a quantum game in which the optimal strategy of three players to win the game corresponds to the case of optimal nonlocality sharing.} 
Our results shed new light on the interplay between nonlocality and quantum measurements, especially the realization of nonlocality sharing via weak measurements, and may find applications in quantum game \cite{Game1, Game2, Game3, Game4}, unbounded randomness certification \cite{Curchod}, quantum coherence \cite{Datta}, quantum steering \cite{Shenoy, Yeon} and quantum communication network \cite{Kimble}.

\section{Methods}

\noindent{\bf{Optical weak measurement in H/V basis}} 
Let us focus attention on Fig.~3(b), and assume that the state after HWP1 and before the first beam displacer (BD) is $\vert\psi\rangle=\alpha\vert H\rangle+\beta\vert V\rangle$. After passing through the first BD, the state may be written as $\alpha\vert H\rangle\vert l\rangle+\beta\vert V\rangle\vert u\rangle$, where $\vert u\rangle$ and $\vert l\rangle$ denote the upper and lower spatial modes, respectively. HWP3 is used to converts $\vert H\rangle$ to $\cos\theta \vert H\rangle+\sin\theta \vert V\rangle$. HWP2 and the $0^\circ$ half-wave plate then convert $\vert V\rangle$ to $\sin\theta \vert H\rangle+\cos\theta \vert V\rangle$. As a result, after passing through HWP2,  the $0^\circ$ half-wave plate, and HWP3, the state becomes  $\alpha(\cos\theta \vert H\rangle+\sin\theta \vert V\rangle)\rangle\vert l\rangle+\beta(\sin\theta \vert H\rangle+\cos\theta \vert V\rangle)\vert u\rangle$. The following three BDs and four $45^\circ$ HWPs realize a swapping operation between the polarization and the spatial degrees of freedom, i.e., $\vert H\rangle\vert l\rangle\rightarrow\vert H\rangle\vert l\rangle$, $\vert H\rangle\vert u\rangle\rightarrow\vert V\rangle\vert l\rangle$, $\vert V\rangle\vert l\rangle\rightarrow\vert H\rangle\vert u\rangle$ and $\vert V\rangle\vert u\rangle\rightarrow\vert V\rangle\vert u\rangle$. As a result, the final output state is given by $\alpha\vert H\rangle(\cos\theta \vert l\rangle+\sin\theta \vert u\rangle)+\beta\vert V\rangle(\sin\theta \vert l\rangle+\cos\theta \vert u\rangle)$. By measuring the spatial qubit, one can realize an optimal weak measurement on the polarization qubit in $H/V$ basis with a precision factor $G=\cos2\theta$.

\section{Acknowledgment}
This work was supported by     National Key R$\&$D Program of China ( 2017YFA0305200,  2016YFA0301300), The Key R$\&$D Program of Guangdong Province ( 2018B030329001, 2018B030325001), The National Natural Science Foundation of China (61974168, 11605205), the Youth Innovation Promotion Association (CAS) (2015317), the Natural Science Foundation of Chongqing (cstc2018jcyjAX0656), the Entrepreneurship and Innovation Support Program for Chongqing Overseas Returnees (cx2017134, cx2018040), the fund of CAS Key Laboratory of Microscale Magnetic Resonance,  and the fund of CAS Key Laboratory of Quantum Information.

%\section{Disclosures}
%The authors declare no conflicts of interest.
%========================================================================================


\begin{thebibliography}{99}

\bibitem{Brunner}N. Brunner, D. Cavalcanti, S. Pironio, V. Scarani, and S. Wehner, Rev. Mod. Phys. \textbf{86}, 419 (2014).
% \bibitem{Einstein} A. Einstein, B. Podolsky, N. Rosen, Phys. Rev. 47, 777 (1935).



\bibitem{Bell} J. S. Bell, Physics (Long Island City, N.Y.) \textbf{1}, 195 (1964).
\bibitem{Clauser} J. Clauser, M. Horne, A. Shimony, R. Holt, Phys. Rev. Lett. \textbf{23}, 880 (1969).
\bibitem{Zukowski} M.\.{Z}ukowski and \v{C}. Brukner, Phys. Rev. Lett. \textbf{88}, 210401 (2002).
\bibitem{Mermin} N. D. Mermin, Phys. Rev. Lett. \textbf{65}, 1838 (1990).
\bibitem{Belinskii} A.V. Belinskii and D. N. Klyshko, Phys. Usp. \textbf{36}, 653 (1993).
\bibitem{Ardehali} M. Ardehali, Phys. Rev. A \textbf{46}, 5375 (1992).
\bibitem{Collins} D. Collins, N. Gisin, N. Linden, S. Massar, S. Popescu, Phys. Rev. Lett. \textbf{88}, 040404 (2002).
\bibitem{Brukner} \v{C}. Brukner, M. \.{Z}ukowski, and A. Zeilinger, Phys. Rev. Lett. \textbf{89}, 197901 (2002).
\bibitem{Lee} S. M. Lee, M. Kim, H. Kim, H. S. Moon, and S. W. Kim, Quantum Science and Technology, \textbf{3}(4), 045006 (2018).

\bibitem{Aspect} A. Aspect, J. Dalibard, and G. Roger, Phys. Rev. Lett. \textbf{49}, 1804 (1982).
\bibitem{Weihs} G. Weihs, T. Jennewein, C. Simon, H. Weinfurter, and A. Zeilinger, Phys. Rev. Lett. \textbf{81}, 5039 (1998).
\bibitem{Rowe} M. A. Rowe, D. Kielpinski, V. Meyer, C. A. Sackett, W. M. Itano, C. Monroe, and D. J. Wineland, Nature \textbf{409}, 791 (2001).
\bibitem{Hofmann} J. Hofmann \emph{et al.}, Science \textbf{337}, 72 (2012).
\bibitem{Giustina} M. Giustina \emph{et al.}, Nature \textbf{497}, 227 (2013).
\bibitem{Christensen} B. G. Christensen \emph{et al.}, Phys. Rev. Lett. \textbf{111}, 130406 (2013).
\bibitem{Hensen} B. Hensen \emph{et al.}, Nature 526, 682 (2015).
\bibitem{Giustina1} M. Giustina, M. A. M. Versteegh, S. Wengerowsky, J. Handsteiner, A. Hochrainer, K. Phelan \emph{et al.}, Phys. Rev. Lett. \textbf{115}, 250401 (2015).
\bibitem{Shalm} L. K. Shalm,E. Meyer-Scott, B. G. Christensen, P. Bierhorst, M. A. Wayne, M. J. Stevens, \emph{et al.}, Phys. Rev. Lett. \textbf{115}, 250402 (2015).


\bibitem{Acin} A. Ac\'{i}n, Phys. Rev. Lett. \textbf{98}, 230501 (2007).

\bibitem{Pironio} S. Pironio \emph{et al.}, Nature \textbf{464}, 1021-1024 (2010).
\bibitem{Liu}Y. Liu, \emph{et al.}, Nature \textbf{562}, 548-551 (2018).
\bibitem{Bierhorst}P. Bierhorst, \emph{et al.}, Nature \textbf{556}, 223-226 (2018).

\bibitem{Bowles} J. Bowles, I. Supic, D. Cavalcanti, A. Acin, Phys. Rev. Lett. \textbf{121}, 180503 (2018).

\bibitem {Silva} R. Silva, N. Gisin, Y. Guryanova, and S. Popescu, Phys. Rev. Lett. \textbf{114}, 250401 (2015).
\bibitem {Hu} M. J. Hu, Z. Y. Zhou, X. M. Hu, C. F. Li, G. C. Guo, and Y. S. Zhang, Npj. Quantum. Inform. \textbf{4}, 63 (2018).
\bibitem {Schiavon} M. Schiavon, L. Calderaro, M. Pittaluga, G. Vallone, and P. Villoresi, Quantum Sci. Technol. \textbf{2}, 015010 (2017).
\bibitem{DAS}D. Das, A. Ghosal, S. Sasmal, S. Mal, and A. S. Majumdar, Physical Review A \textbf{99}, 022305 (2019).
\bibitem{Mal}S. Mal, A. S. Majumdar, and D. Home, Mathematics, \textbf{4}(3), 48, (2016).
\bibitem{Ren}C. Ren, T. Feng, D. Yao, H. Shi, J. Chen, and X. Zhou, Physical Review A, \textbf{100}(5), 052121(2019).

\bibitem{Masanes} L. Masanes, A. Acin, and N. Gisin, Phys. Rev. A \textbf{73}, 012112 (2006).
\bibitem{Toner} B. Toner, Proceedings of the Royal Society A: Mathematical, Physical and Engineering Sciences \textbf{465}, 59 (2009).

\bibitem{Game1} N. D. Mermin, Am. J. Phys. \textbf{58}, 731 (1990).
\bibitem{Game2} N. D. Mermin, Phys. Rev. Lett. \textbf{65}, 3373 (1990)
\bibitem{Game3}  A. Peres, Phys. Lett. A \textbf{151}, 107 (1990)
\bibitem{Game4} N. Brunner and N. Linden, Nat. Commun. \textbf{4}, 2057 (2013).


\bibitem{Curchod} F. J. Curchod, M. Johansson, R. Augusiak, M. J. Hoban, P. Wittek, A. Acin, Phys. Rev. A \textbf{95}, 020102(R) (2017).

\bibitem{Datta} S. Datta and A. S. Majumdar, Phys. Rev. A \textbf{98}, 042311 (2018).
\bibitem{Shenoy} A. Shenoy H., S. Designolle, F. Hirsch, R. Silva, N. Gisin, N. Brunner, Phys. Rev. A \textbf{99}, 022317 (2019).
\bibitem{Yeon} Y. Choi, S. Hong, T. Pramanik, H. Lim, Y. Kim, H. Jung, S. Han, S. Moon, and Y. Cho, Optica \textbf{7}(6), 675 (2020)

\bibitem{Kimble}H. J. Kimble, Nature \textbf{453}, 1023-1030 (2008).


%\bibitem{Acin} A. Ac\'{i}n, Phys. Rev. Lett. \textbf{98}, 230501 (2007).
%
%\bibitem{Pironio} S. Pironio et al., Nature \textbf{464}, 1021-1024 (2010).
%\bibitem{Liu}Y. Liu, et al.,  Nature \textbf{562}, 548-551 (2018).
%\bibitem{Bierhorst}P. Bierhorst,  et al., Nature \textbf{556}, 223-226 (2018).
%
%\bibitem{Bowles} J. Bowles, I. Supic, D. Cavalcanti, A. Acin, Phys. Rev. Lett. \textbf{121}, 180503 (2018).
%
%\bibitem {Silva} R. Silva, N. Gisin, Y. Guryanova, and S. Popescu, Phys. Rev. Lett. \textbf{114}, 250401 (2015).
%\bibitem {Hu} M. J. Hu, Z. Y. Zhou, X. M. Hu, C. F. Li, G. C. Guo, and Y. S. Zhang, Npj. Quantum. Inform. \textbf{4}, 63 (2018).
%\bibitem {Schiavon} M. Schiavon, L. Calderaro, M. Pittaluga, G. Vallone, and P. Villoresi, Quantum Sci. Technol. \textbf{2}, 015010 (2017).
%
%\bibitem{Ren}C. Ren, T.  Feng, D. Yao,  H. Shi,  J.  Chen, and X. Zhou, Physical Review A, 100(5), 052121(2019).
%
%\bibitem{Masanes} L. Masanes, A. Acin, and N. Gisin, Phys. Rev. A \textbf{73}, 012112 (2006).
%\bibitem{Toner} B. Toner, Proceedings of the Royal Society A: Mathematical, Physical and Engineering Sciences \textbf{465}, 59 (2009).
%
%\bibitem{Curchod} F. J. Curchod, M. Johansson, R. Augusiak, M. J. Hoban, P. Wittek, A. Ac\'{i}n, Phys. Rev. A \textbf{95}, 020102(R) (2017).
%
%\bibitem{Datta} S. Datta and A. S. Majumdar, Phys. Rev. A \textbf{98}, 042311 (2018).
%%\bibitem{Shenoy} A. Shenoy, et al.,  Phys. Rev. A \textbf{99}, 022317 (2019).
%\bibitem{Shenoy} A. Shenoy H.,  S. Designolle, F. Hirsch, R. Silva, N. Gisin, N. Brunner,  Phys. Rev. A \textbf{99}, 022317 (2019).
%%\bibitem{Shenoy} A. Shenoy H., S. Designolle, F. Hirsch, R. Silva, N. Gisin, N. Brunner,  Phys. Rev. A \textbf{99}, 022317 (2019).
%
%\bibitem{Kimble}H. J. Kimble,  Nature \textbf{453}, 1023-1030 (2018).




\end{thebibliography}
\end{document}